\pdfoutput=1
\documentclass[aps,prc,twocolumn,superscriptaddress,showpacs]{revtex4}

\usepackage{graphicx}
\usepackage{dcolumn}
\usepackage{bm}
\usepackage{ulem}
\usepackage{color}

\newcommand{\al}{$\alpha$}
\newcommand{\g}{$\gamma$}

\newcommand{\raX}{($\alpha$,$X$)}
\newcommand{\raapr}{($\alpha$,$\alpha'$)}
\newcommand{\rag}{($\alpha$,$\gamma$)}
\newcommand{\ran}{($\alpha$,n)}
\newcommand{\raan}{($\alpha$,$\alpha$n)}

\newcommand{\rann}{($\alpha$,2n)}
\newcommand{\rannn}{($\alpha$,3n)}

\newcommand{\rap}{($\alpha$,p)}
\newcommand{\rapn}{($\alpha$,pn)}

\newcommand{\rga}{($\gamma$,$\alpha$)}

\newcommand{\stot}{$\sigma_{\rm{reac}}$}

\newcommand{\zniv}{$^{64}$Zn}

\newcommand{\gevii}{$^{67}$Ge}
\newcommand{\geviii}{$^{68}$Ge}
\newcommand{\gavii}{$^{67}$Ga}

\newcommand{\ppro}{p-process}
\newcommand{\gpro}{$\gamma$-process}
\newcommand{\pnuc}{p nucleus}
\newcommand{\sfact}{S-factor}
\newcommand{\Nsv}{$N_A$$\left< \sigma v \right>$}
\newcommand{\fdev}{$f_{\rm{dev}}$}
\newcommand{\Ndgg}{$N_{\rm{DGG}}$}

\begin{document}
\title{
Statistical model analysis of $\alpha$-induced reaction cross sections of
$^{64}$Zn at low energies
}
\author{P.~Mohr}
\email{mohr@atomki.mta.hu}
\affiliation{
Institute for Nuclear Research (MTA Atomki), H-4001 Debrecen, Hungary}
\affiliation{
Diakonie-Klinikum, D-74523 Schw\"abisch Hall, Germany}
\author{Gy.~Gy\"urky}
\affiliation{
Institute for Nuclear Research (MTA Atomki), H-4001 Debrecen, Hungary}
\author{Zs.~F\"ul\"op}
\affiliation{
Institute for Nuclear Research (MTA Atomki), H-4001 Debrecen, Hungary}
\date{\today}

\begin{abstract}
\begin{description}
\item[Background] $\alpha$-nucleus potentials play an essential role for the
  calculation of $\alpha$-induced reaction cross sections at low energies in
  the statistical model. Uncertainties of these calculations are related to
  ambiguities in the adjustment of the potential parameters to experimental
  elastic scattering angular distributions (typically at higher energies) and
  to the energy dependence of the effective $\alpha$-nucleus potentials.
\item[Purpose] The present work studies cross sections of
  $\alpha$-induced reactions for $^{64}$Zn at low energies and their
  dependence on the chosen input parameters of the statistical model
  calculations. The new experimental data from the recent Atomki experiments
  allow for a $\chi^2$-based estimate of the uncertainties of calculated cross
  sections at very low energies.
\item[Method] The recently measured data for the ($\alpha$,$\gamma$),
  ($\alpha$,$n$), and ($\alpha$,$p$) reactions on $^{64}$Zn are compared to
  calculations in the statistical model. A survey of the parameter space of
  the widely used computer code TALYS is given, and the properties of
  the obtained $\chi^2$ landscape are discussed.
\item[Results] The best fit to the experimental data at low energies shows
  $\chi^2/F \approx 7.7$ per data point which corresponds to an average
  deviation of about 30\,\% between the best fit and the experimental
  data. Several combinations of the various ingredients of the statistical
  model are able to reach a reasonably small $\chi^2/F$, not exceeding the
  best-fit result by more than a factor of two.
\item[Conclusions] The present experimental data for $^{64}$Zn in combination
  with the statistical model calculations allow to constrain the astrophysical
  reaction rate within about a factor of 2. However, the significant excess of
  $\chi^2/F$ of the best-fit from unity asks for further improvement
  of the statistical model calculations and in particular the $\alpha$-nucleus
  potential.
\end{description}
\end{abstract}

\maketitle

\section{Introduction}
\label{sec:intro}
Statistical model calculations are widely used for the calculation of reaction
cross sections of \al -induced reactions for intermediate mass and heavy
target nuclei. It is found that the cross sections at low energies depend
sensitively on the chosen \al -nucleus optical model potential. These
low-energy cross sections play also an essential role under stellar
conditions. In particular, \rga\ photodisintegration reactions in the
so-called astrophysical \ppro\ (or \gpro )
are best determined by the study of the inverse
\rag\ capture reactions \cite{Mohr07}. Under typical \ppro\ conditions,
temperatures of about 
$2-3$ billion Kelvin are reached (in short: $T_9 = 2-3$), and the
corresponding Gamow window is located at a few MeV for intermediate mass
nuclei like \zniv\ (e.g.\ \cite{Woo78,Arn03,Rau06,Rap06,Tra14}).

The lightest nucleus which is synthesized in the \ppro , is $^{74}$Se
\cite{Woo78}. These so-called p nuclei are typically characterized by very low
natural abundances of the order of 1\,\% or even below. Although \zniv\ is
somewhat lighter than the lightest \pnuc , it has nevertheless been chosen for
the present study because the cross sections of various \al -induced reactions
can be determined for \zniv\ by the simple and robust activation technique
with high precision. The high natural abundance of \zniv\ of almost 50\,\%
allows to use targets with natural isotopic composition. However, there is
also a drawback. For lighter nuclei like \zniv\ the applicability of the
statistical model may be limited at very low energies because of an
insufficient level density in the \geviii\ compound nucleus.

Our recent study of \al -induced reaction cross sections for the target
nucleus \zniv\ \cite{Orn16,Gyu12} was focused on total reaction cross sections
\stot\ and its determination from either elastic scattering angular
distributions or from the sum over the cross sections of all open non-elastic
channels (including inelastic scattering). It was found that there is
excellent agreement at the lower energy of 12.1\,MeV ($428 \pm 7$\,mb from
elastic scattering vs.\ $447 \pm 41$\,mb from the sum over non-elastic
channels). At the higher energy of 16.1\,MeV a significant contribution of
compound-inelastic \raapr\ scattering to higher-lying states in \zniv\ was
identified which is about $15-20$\,\% of \stot $= 905 \pm 18$\,mb from elastic
scattering.

The present study extends our previous work \cite{Orn16,Gyu12} by a detailed
study of the \rag , \ran , \rap , and total reaction cross sections and their
dependence on the underlying parameters of the statistical model (StM)
calculations. For this purpose the full parameter space of the widely used
TALYS \cite{TALYS} code (version 1.8) was investigated. In particular, all
available options for the \al -nucleus optical model potential (A-OMP), the
nucleon OMP (N-OMP), the \g -ray strength function (GSF), and the level
density (LD) were varied. Almost 7,000 combinations of input parameters are
used to calculate a $\chi^2$ landscape. This landscape provides improved
insight into the sensitivities of the different reaction channels on the
underlying parameters. It is the scope of the present study to obtain an
improved $\chi^2$-based prediction of reaction cross sections at very low
energies where experimental data are not available. As an example, an
extrapolation to the astrophysically most relevant energies is made for the
\zniv \rag \geviii\ reaction with an estimate of the corresponding
uncertainties. 

The most important ingredient for the calculation of \al -induced reaction
cross sections in the StM is the A-OMP. For heavy nuclei (above $A \gtrsim
100$) it has been found that different A-OMPs lead to dramatic variations of
the predicted cross sections, exceeding one order of magnitude at
astrophysically relevant energies (e.g., \cite{Som98,Sch16}). Contrary, a
quite reasonable description of the recent data for \zniv\ was found for
several A-OMPs \cite{Orn16}. However, as will be shown below from a
$\chi^2$-based assessment, this reasonable description for \zniv\ is still far
from a precise prediction of the experimental results.

The paper is organized as follows. In Sect.~\ref{sec:para} a brief description
of the statistical model and its ingredients is given. Available experimental
data are summarized in Sect.~\ref{sec:exp}. Sect.~\ref{sec:res} presents the
results and shows the obtained excitation functions for the total
reaction cross section \stot\ and the \rag , \ran , and \rap\ reaction
channels under study. A discussion of the results and an extrapolation to
lower energies is provided in Sect.~\ref{sec:disc}. Conclusions are drawn in
Sect.~\ref{sec:conc}.

\section{Parameters of the Statistical Model}
\label{sec:para}

\subsection{Basic Considerations}
\label{sec:basic}
In a schematic notation the reaction cross section in the StM is proportional
to 
\begin{equation}
\sigma(\alpha,X) \sim \frac{T_{\alpha,0} T_X}{\sum_i T_i} = T_{\alpha,0}
\times b_X
\label{eq:StM}
\end{equation}
with the transmission coefficients $T_i$ into the $i$-th open channel and the
branching ratio $b_X = T_X / \sum_i T_i$ for the decay into the channel
$X$. The total transmission is given by the sum over all contributing
channels: $T_{\rm{tot}} = \sum_i T_i$. The $T_i$ are calculated from global
optical potentials (A-OMP and N-OMP for the particle channels) and from the
GSF for the photon channel. The $T_i$ include contributions of all final
states $j$ in the respective residual nucleus in the $i$-th exit channel. In
practice, the sum over all final states $j$ is approximated by the sum over
low-lying excited states up to a certain excitation energy $E_{\rm{LD}}$
(these levels are typically known from experiment) plus an integration over
a theoretical level density for the contribution of higher-lying excited states:
\begin{equation}
T_i = \sum_j T_{i,j} \approx 
\sum_j^{E_j < E_{\rm{LD}}} T_{i,j} +
\int_{E_{\rm{LD}}}^{E_{\rm{max}}} \rho(E) \, T_i(E) \, dE
\label{eq:Tsum}
\end{equation}
For further details of the definition of $T_i$, see
\cite{Rau11}. $T_{\alpha,0}$ refers to the entrance channel where the target
nucleus is in its ground state under laboratory conditions. The calculation of
stellar reaction rates \Nsv\ requires further modifications of
Eq.~(\ref{eq:StM}) which have to take into account thermal excitations of the
target nucleus \cite{Rau11}. For the \rag\ reaction of the even-even
nucleus \zniv\ with the relatively high excitation energy of the first excited
state ($J^\pi = 2^+$, $E^\ast = 992$\,keV), these corrections remain small at
typical temperatures of the \ppro\ of the order of a few billion Kelvin
\cite{Rau00,Rau11a}.

The properties of the $T_i$ in Eqs.~(\ref{eq:StM}) and (\ref{eq:Tsum}) lead to
the following general findings for the case of \zniv . At very low energies,
the \rag\ channel with its positive $Q$-value of $Q_\gamma = +3.40$\,MeV is
the only open reaction channel besides elastic and inelastic scattering. The
transmission $T_\gamma$ into the \g -channel exceeds the transmission
$T_\alpha$ into the \al -channel which is strongly suppressed by the Coulomb
barrier. Consequently, $\sum_i T_i \approx T_\gamma$, and the \rag\ cross
section becomes proportional to $T_{\alpha,0}$, but almost independent of the
other $T_i$ (including $T_\gamma$).

As soon as the proton channel opens ($Q_p = -3.99$\,MeV), $T_p$ increases
almost exponentially with energy and exceeds $T_\gamma$ already close above
the proton threshold. Because of the lower Coulomb barrier for the
proton channel (compared to the \al\ case), $T_p$ becomes the dominant
contributor to the sum in $T_{\rm{tot}}$. This leads to a \rap\ cross section
proportional to $T_{\alpha,0}$ but independent of the other $T_i$ (including
$T_p$). The \rag\ cross section becomes approximately proportional to
$T_{\alpha,0} T_\gamma / T_p$.

Because of the strongly negative $Q$-value of the \ran\ channel ($Q_n =
-8.99$\,MeV) and the resulting smaller phase space (in comparison to the
\rap\ reaction), the contribution of the \ran\ channel remains relatively
small. This finding is different from heavy nuclei where the \ran\ channel
typically becomes dominant close above the neutron threshold (see
e.g.\ \cite{Mohr11}). Now we find the approximate proportionalities of
$T_{\alpha,0} T_p / (T_p + T_n)$ for the \rap\ cross section, $T_{\alpha,0}
T_n / (T_p + T_n)$ for the \ran\ cross section, and $T_{\alpha,0} T_\gamma /
(T_p + T_n)$ for the \rag\ cross section.

Although the above discussion is indeed somewhat simplistic, it is
nevertheless helpful for a general understanding of the sensitivities of the
\rag , \ran , and \rap\ cross sections on the underlying
parameters. Sensitivities as a function of energy have been calculated by
Rauscher \cite{Rau12} for a wide range of nuclei; the results from the
NON-SMOKER code are available online at \cite{NONSMOKER} and confirm the above
discussion for \zniv .

The role of the level density requires further discussion. As
Eq.~(\ref{eq:Tsum}) shows, the chosen parametrization of the LD becomes only
relevant above a certain number of low-lying excited states which are taken
into account explicitly. Typically, these low-lying levels cover an excitation
energy range of $E^\ast \approx 2 - 3$\,MeV. Thus, for \al -induced reactions
on \zniv\ this means that the importance of the LDs remains limited, in
particular at low energies, for the \ran\ and \rap\ reactions whereas the role
of the LD is significant for the \rag\ reaction; here the last term in
Eq.~(\ref{eq:Tsum}) does contribute.

However, besides its relatively minor role for the calculation of the
transmissions $T_i$ in Eq.~(\ref{eq:Tsum}), the LD plays an essential role for
the applicability of the StM which is only valid if the experimental
conditions (mainly target thickness and energy distribution of the beam)
average over a sufficient number of levels in the compound nucleus \geviii
. For the present data the energy interval $\Delta E$ of the experiment is
of the order of 100\,keV \cite{Gyu12}. The experimental energies cover an
energy range of about 6\,MeV $\le E_{\rm{c.m.}} \le$ 15\,MeV. Together with
the $Q$-value of the \rag\ reaction of about $+4$\,MeV, this corresponds to
excitation energies $E^\ast$ in \geviii\ of about 10\,MeV $\le E^\ast \le
19$\,MeV. 

The various options for the LD in TALYS (see Sect.~\ref{sec:LD}) predict total
level densities (per parity) between about 9000\,MeV$^{-1}$ and
80000\,MeV$^{-1}$ already at the lowest energies under study. At first view,
this seems to be sufficient for the applicability of the StM. But in
particular for the \rap\ reaction close above the threshold, the experimental
excitation curve is not completely smooth (as expected for a fully statistical
behavior). Close above the threshold, the \zniv \rap \gavii\ reaction
populates only very few final states in \gavii\ with low spins $J$ and
negative parity. The barrier penetration in this exit channel favors proton
emission with low angular momentum, and thus by far not all levels in the
\geviii\ compound nucleus contribute to the \zniv \rap \gavii\ cross
section. A closer look at the level densities in \geviii\ shows that the
predicted level density per spin goes down to e.g.\ about a few hundred per
MeV for $J = 0$, or less than 100 levels may contribute to the averaging
within the experimental energy interval of $\Delta E \approx 100$\,keV. Thus,
non-statistical fluctuations in the excitation function of \zniv \rap
\gavii\ at low energies are not very surprising.

LDs increase dramatically with increasing excitation energy. At the highest
energies under study, the predicted level densities are at least two orders of
magnitude higher than at the lowest energies. And indeed the non-statistical
fluctuations in the excitation function of \zniv \rap \gavii\ disappear at
energies above 10\,MeV (corresponding to $E^\ast \approx 14$\,MeV in
\geviii\ or an increase of the level density by more than one order of
magnitude, compared to $E^\ast \approx 10$\,MeV).

\subsection{Ingredients under Consideration}
\label{sec:ingr}

\subsubsection{\al -nucleus optical model potentials}
\label{sec:AOMP}
The \al -nucleus optical model potential is the essential ingredient for the
calculation of \al -induced reaction cross sections at low energies. TALYS
provides 8 options for the A-OMP: The first option is based on the early work of
Watanabe \cite{Wat58}; this was the default option in TALYS. The widely used
simple 4-parameter potential by McFadden and Satchler \cite{McF66} is the second
option in TALYS. Three versions of the double-folding A-OMP by Demetriou {\it
  et al.}\ \cite{Dem02} (DGG-1, DGG-2, DGG-3) are also included in
TALYS. Since TALYS version 1.8, three additional A-OMPs are available
which are based on Nolte {\it el al.}\ \cite{Nol87} and on two versions of
Avrigeanu {\it et al.}\ \cite{Avr14,Avr94} (AVR for the latest version in
\cite{Avr14}).

In addition, the new ATOMKI-V1 potential \cite{Mohr13} was implemented into
the TALYS code, and modifications of the third version of the Demetriou
potential DGG-3 were investigated. It was recently suggested in \cite{Sch16}
that the real part of this potential should be multiplied by a factor of about
\Ndgg\ $\approx 1.1 - 1.2$ for a better description of reaction data for heavy
targets ($A \gtrsim 100$). For $^{64}$Zn it will be shown that the best fit is
achieved by a reduction of the real part by a factor of about 0.9 (instead of
an increased potential as found for $A \gtrsim 100$ in \cite{Sch16}).

The latest global A-OMP by Su and Han \cite{Su15} is not yet implemented in
TALYS. It has been shown in \cite{Orn16} that this potential overestimates the
total reaction cross sections for \zniv\ at low energies. Thus, no efforts
have been made to implement this potential into TALYS for the present study.

\subsubsection{Nucleon-nucleus optical model potentials}
\label{sec:NOMP}
The default option in TALYS is based on the local and global parametrizations
in Koning and Delaroche \cite{Kon03}. Furthermore, based on the work of
Jeukenne, Lejeune, and Mahaux (e.g., \cite{Jeu77}), four different versions of
the so-called JLM potential are available. The basic JLM-type potential is
taken from Bauge {\it et al.}\ \cite{Bau01}, and three modifications of this
potential are taken from Goriely and Delaroche \cite{Gor07}.

\subsubsection{\g -ray strength functions}
\label{sec:GSF}
Eight different options for the \g -ray strength function are implemented in
TALYS. In general, the options are based on the work of Brink \cite{Bri57} and
Axel \cite{Axel62} or Kopecky and Uhl \cite{Kop90}. In addition, microscopic
model GSFs have been calculated on the basis of Hartree-Fock BCS,
Hartree-Fock-Bogolyubov, relativistic mean field models \cite{Cap09}, and a
hybrid model \cite{Gor98}. Furthermore, the above choices can be combined with
two options for the M1 strength function where the M1 strength is either
normalized to the E1 strength (default option) or not normalized. A detailed
description of the available options can be found in the TALYS manual
\cite{TALYS} and in the Reference Input Parameter Library (RIPL) \cite{Cap09}.

\subsubsection{Level densities}
\label{sec:LD}
Three phenomenological and three miscroscopic level densities can be chosen in
TALYS. The phenomenological options are based on constant temperature and
back-shifted Fermi gas models and on the generalized superfluid model. The
microscopic approaches are calculated using Skyrme or Gogny type
forces. Details on the various options are summarized in \cite{Kon08}.

\section{Experimental data}
\label{sec:exp}
\begin{figure*}[htb]
\includegraphics[bbllx=35,bblly=20,bburx=790,bbury=595,width=17.5cm,clip=]{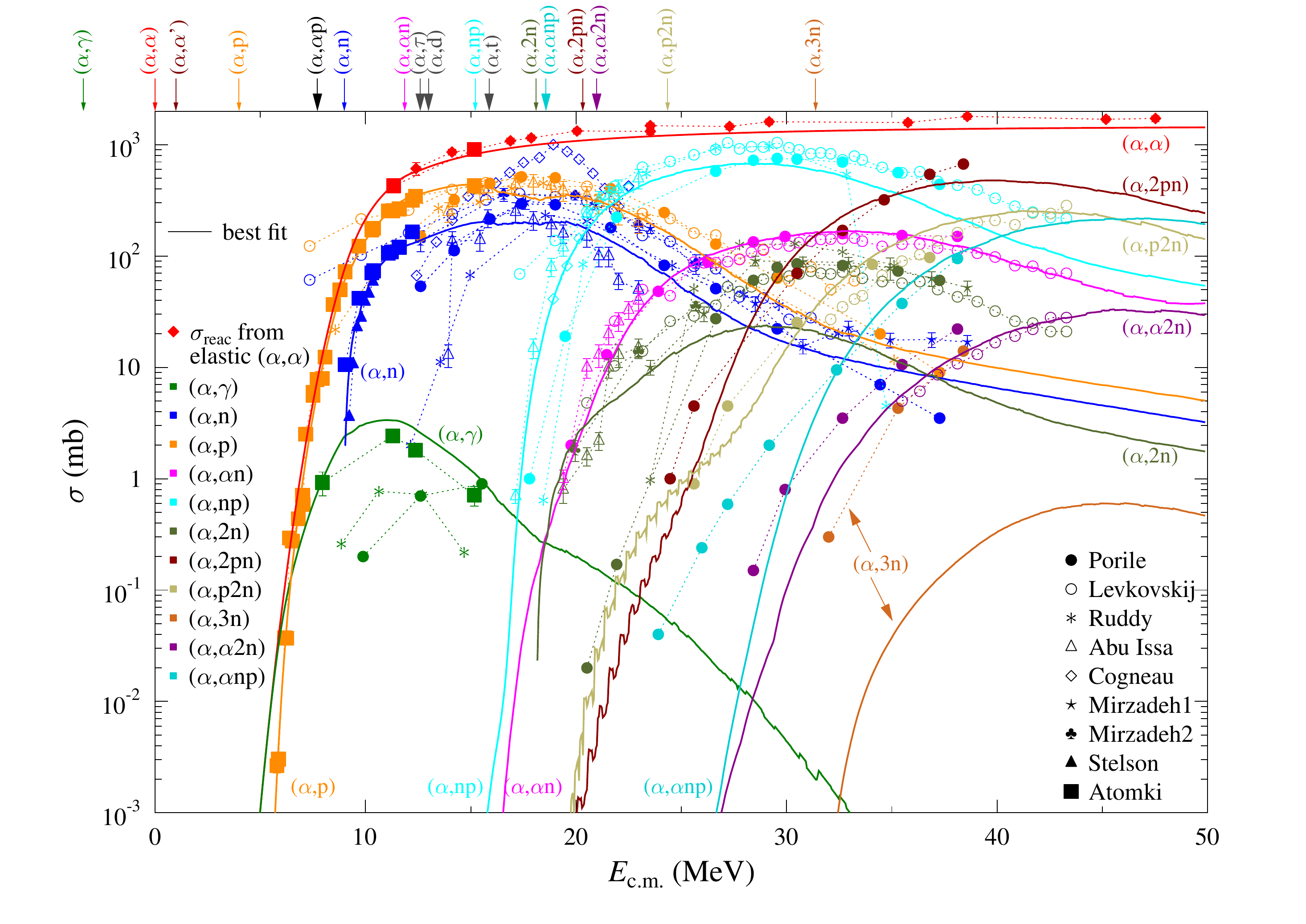}
\caption{
\label{fig:reac_all}
(Color online)
Cross sections of \al -induced reactions on \zniv\ over a wide energy
range. The data labeled \stot\ (shown in red) are total reaction cross sections
and have been derived from elastic scattering angular distributions
\cite{Orn16}. The thresholds for the different channels are indicated by arrows
on top. Data sets for different reaction channels are shown in different
colors; the \rag , \ran , and \rap\ reactions are shown in green, blue, and
orange. Different symbols represent the different data sets by Porile
\cite{Por59}, Levkovskij \cite{Lev91}, Ruddy {\it et al.}\ \cite{Rud64}, Abu
Issa {\it et al.}\ \cite{Abu89}, Cogneau {\it et al.}\ \cite{Cog65}, two data
sets by Mirzadeh {\it et al.}\ \cite{Mir91}, Stelson {\it et
  al.}\ \cite{Ste64}, and the recent Atomki data \cite{Orn16,Gyu12}. All data
sets are connected by thin dotted lines to guide the eye. The full lines are
the best-fit TALYS calculations (see Sec.~\ref{sec:res}). Further discussion
see text.
}
\end{figure*}

Several excitation functions of \al -induced reaction cross sections for
\zniv\ are available in literature and in the EXFOR database
\cite{EXFOR}. However, either the data are more than 50 years old
\cite{Por59,Rud64,Ste64,Cog65}, or the data have not been published in
refereed journals \cite{Abu89,Mir91,Lev91}. All EXFOR data are presented in
Fig.~\ref{fig:reac_all}. Unfortunately, significant discrepancies between the
different excitation functions are found (see Fig.~\ref{fig:reac_all}). 

Although there is no good agreement between the different experimental data
sets, Fig.~\ref{fig:reac_all} nicely shows that the overall energy dependence
of the various reaction channels is well reproduced by the best-fit
statistical model calculation. The only exceptions are the \rannn\ reaction
where the only available experimental data set of Porile \cite{Por59} is more
than one order of magnitude above the theoretical estimate, and the
\rann\ reaction which is underestimated by about a factor of 4. 

Furthermore, the importance of the various reaction channels at different
energies can nicely be read from Fig.~\ref{fig:reac_all}. At very low energies
below about 6\,MeV the \rag\ reaction is dominating because the \rap\ and
\ran\ channels are closed or suppressed by the Coulomb barrier of the outgoing
low-energy proton. At about 6\,MeV the \rap\ reaction starts to dominate up to
almost 20\,MeV. As soon as the \ran\ channel opens, also a significant
contribution of the \ran\ channel is found. Above 20\,MeV, various
multi-particle emission channels like \rapn , \rann , and \raan\ contribute
also to the total reaction cross section \stot .

As the focus of the present study is the low-energy region, we finally decided
to use only our latest data of the \al -induced cross sections for \zniv\ for
the determination of the best-fit parameters for the statistical model
calculations at low energies. In particular, this means 4 data points for the
\rag\ reaction, 10 data points for the \ran\ reaction, 27 data points for the
\rap\ reaction, and 2 data points for the total reaction cross section
\stot\ from the analysis of the elastic scattering angular distributions
\cite{Orn16,Gyu12}; in total, 43 experimental data points are used to
determine the $\chi^2$ landscape.  The $\chi^2$ adjustment procedure will be
discussed in detail in the following section.

\section{Results for $\chi^2/F$}
\label{sec:res}
All combinations of the A-OMPs, N-OMPs, GSFs, and LD parametrizations in
Sect.~\ref{sec:ingr} have been used for the calculation of the \rag , \ran ,
and \rap\ cross sections of \zniv . In detail this means that 6720
combinations of A-OMPs, N-OMPs, GSFs, and LDs have been calculated. This
number results from 14 A-OMPs (8 built-in in TALYS plus ATOMKI-V1 plus DGG-3
multiplied by factors of \Ndgg\ $= 0.7$, 0.8, 0.9, 1.1, and 1.2), 8 E1 GSFs
combined with two additional options for the M1 strength, 5 N-OMPs, and 6 LDs.

Technically it would be possible to further increase this number by choosing
different models for each residual nucleus, e.g.\ different N-OMPs for the
neutron and the proton channel or different GSFs or LDs for even and odd
residual nuclei, etc.\ etc. However, best-fit parameters should be valid for a
reasonable range of nuclei, and thus the present work intentionally remained
restricted to the above listed 6720 combinations of A-OMPs, N-OMPs, GSFs, and
LDs, and did not apply different parameter sets for different residual nuclei of
the \al\ + \zniv\ reactions.

%
%
For each of the 6720 combinations of parameters, excitation functions for the
total reaction cross section \stot\ and the cross sections of the \rag , \ran
, and \rap\ reaction channels were calculated, and the deviation between the
theoretical excitation functions and the experimental data was determined by a
standard $\chi^2$ calculation. 

\subsection{$\chi^2/F$ from all experimental data points}
\label{sec:chi2all}
\begin{figure}[htb]
\includegraphics[width=0.82\columnwidth,clip=]{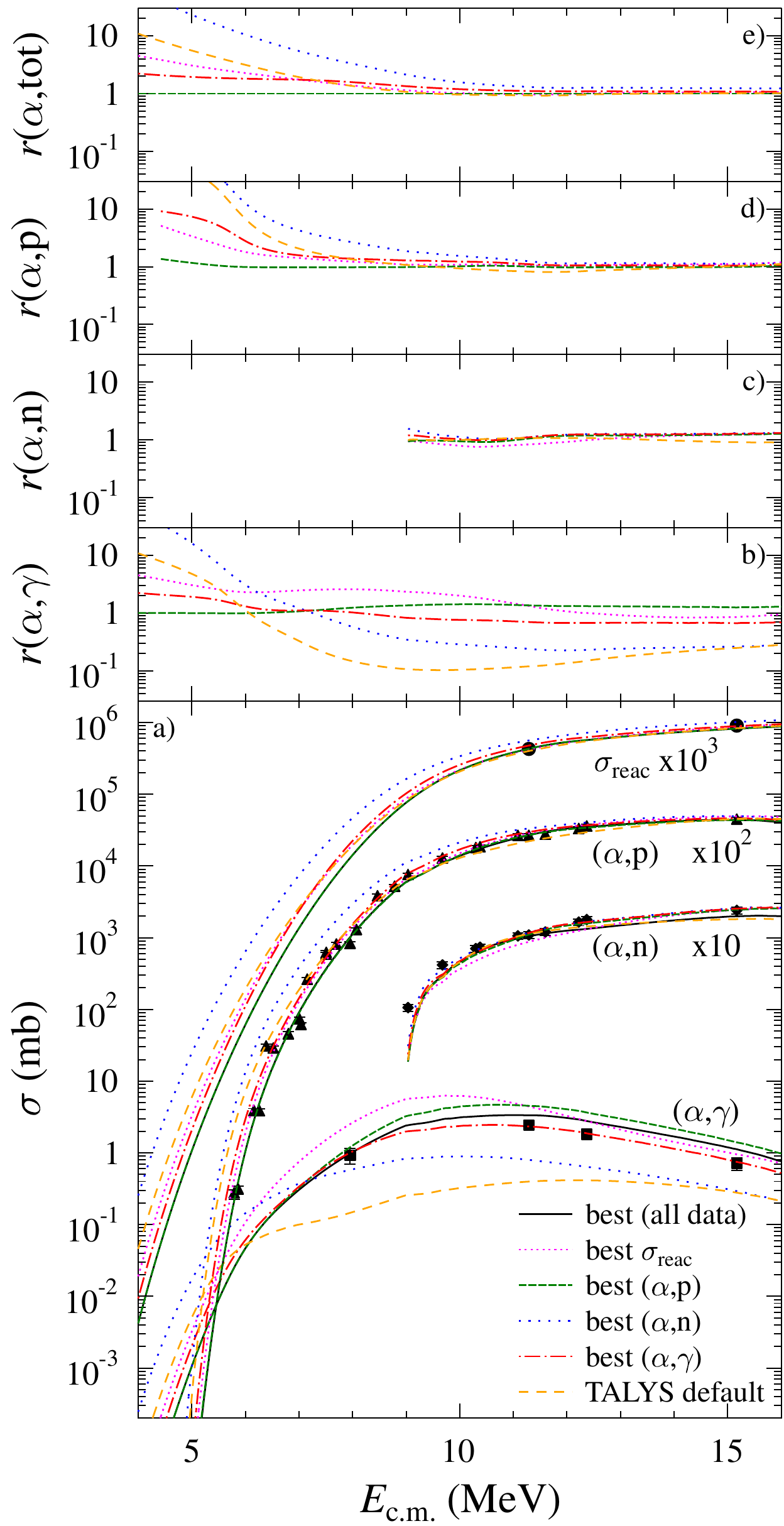}
\caption{
\label{fig:reac_chi2}
(Color online)
Cross sections of \al -induced reactions on \zniv ; for better readability,
the total reaction cross sections \stot\ have been multiplied by a factor of
1000, the \rap\ data by a factor of 100, and the
\ran\ data by a factor of 10. The best fit of all 6720
combinations of the TALYS parameters is shown with a full black line. The best
fits to the \stot\ data and to individual \rap , \ran , and \rag\ reactions are
shown with narrow-dotted magenta, dashed green, dotted blue, and dash-dotted
red lines. The TALYS default calculation is shown with a short-dashed orange
line. The lowest part a) shows the calculated excitation functions in
comparison to the experimental data. For better visualization, the upper parts
b), c), d), e) show the same calculations as a ratio $r$ for a particular
\raX\ channel, normalized to the overall best-fit calculation. E.g., part b)
shows the \rag\ cross sections from the best-fits to the \rag , \ran , \rap ,
and \stot\ data, normalized to the overall best-fit. Further discussion see
text.
}
\end{figure}

It is found that the 6720 combinations under
study show a wide range of $\chi^2/F$ per data point from slightly below 8 to
more than 4000 for all 43 experimental values for \stot\ and the \ran , \rag ,
and \rap\ reactions in \cite{Orn16,Gyu12}. These $\chi^2/F$ correspond to an
average deviation factor \fdev\ between experiment and theory for all data
points from about 1.3 for the best fits up to 2.6 for the worst cases.
Fig.~\ref{fig:reac_chi2} shows the results of the above calculations. The
smallest $\chi^2/F \approx 7.7$ per point was found for the following
combination: the A-OMP was derived from the DGG-3 potential with the real part
multiplied by a factor of \Ndgg\ $=0.9$; the N-OMP is taken from Koning and
Delaroche (TALYS default); the GSF is calculated using the Brink-Axel
Lorentzian model with unnormalized M1 strength; the LD was taken from the
back-shifted Fermi gas model. The obtained $\chi^2/F \approx 7.7$ corresponds
to \fdev\ $\approx 1.3$. Obviously, a $\chi^2/F \approx 7.7$ corresponds on
average to almost a $3 \sigma$ deviation for each data point which is not
fully satisfying. This finding will be discussed later (see
Sect.~\ref{sec:disc}). Furthermore, it has to be pointed out that this
best-fit is indeed restricted to the low-energy data. At higher energies above
about 15\,MeV the DGG-3 A-OMP underestimates the total reaction cross sections
\stot\ significantly, and this deviation increases with decreasing
normalization factor \Ndgg\ (\Ndgg\ $= 0.9$ for the best-fit); see also
Fig.~\ref{fig:reac_all}.

The 4 experimental data points of the \rag\ reaction are reproduced with
$\chi^2/F$ from 0.1 up to more than 1200, corresponding to \fdev\ $\approx
1.05 - 8$. The $\chi^2/F$ for the 10 \ran\ data points show a much smaller
variation of $\chi^2/F$ from 4.7 to 16.3, corresponding to \fdev\ between 1.2
and 1.9. The 27 data points for the \rap\ reaction show a wide variation of
$\chi^2/F$ from 6.6 to 6500, corresponding to \fdev\ between 1.25 and
3.2. Finally, because of the small experimental uncertainties, the total
reaction cross sections \stot\ show significant $\chi^2/F$ between 2.9 and 200
although \fdev\ remains relatively close to unity between 1.03 and 1.21. For
completeness it has to be mentioned that the total reaction cross section
\stot\ is sensitive only to the A-OMP, but independent of the other
ingredients of the StM calculations.

\subsection{$\chi^2/F$ from the individual reaction channels}
\label{sec:chi2indiv}
In addition, Fig.~\ref{fig:reac_chi2} shows also the best fits to the
particular \rap , \ran , and \rag\ channels. Obviously, as soon as the fit is
restricted to a particular \raX\ reaction, the resulting parameters are
different. This becomes e.g.\ very prominent for the \rag\ reaction which
depends sensitively on the combination of the transmission coefficients of
different \al , n, p, and \g\ channels. 

The cross sections in Fig.~\ref{fig:reac_chi2} show a strong energy
dependence, and they cover several orders of magnitude. Thus, for better
visualization we have also included the ratios $r$ between the individual fits
and the best-fit for each of the \rag , \ran , and \rap\ channels in the upper
parts b), c), d) and e) of Fig.~\ref{fig:reac_chi2}. Part c) nicely shows the
minor sensitivity of the \ran\ cross section. From part d) it is obvious that
the dominating \rap\ cross section shows a strong sensitivity to the chosen
parameters at low energies, whereas part b) shows that the \rag\ cross section
varies widely over the whole energy range under study. Consequently, improved
constraints for the StM parameters could be obtained from \rap\ data towards
lower energies (down to threshold) and from \rag\ data with smaller
uncertainties in the full energy range.

\subsubsection{$\chi^2/F$ from \rag\ data}
\label{sec:chi2ag}
The lowest $\chi^2/F \approx 0.13$ for the \rag\ channel is obtained for the
combination of A-OMP: DGG-3 with \Ndgg\ $=  1.2$; N-OMP: Koning and Delaroche
(TALYS default); GSF: hybrid model, M1 strength not normalized; LD: constant
temperature Fermi gas (TALYS default). However, the best-fit parameters of the
\rag\ channel lead to an increased $\chi^2/F$ for all \stot , \rap , \ran ,
and \rag\ data by more than a factor of five to about 40 (compared to 7.7 for
the overall best-fit), and in particular the \rap\ cross section at low
energies is about a factor of 10 higher than the overall best-fit, see 
Fig.~\ref{fig:reac_chi2}d). Here it becomes evident that a restricted analysis
of the \rag\ channel only may be misleading. The \rag\ cross section is
sensitive to the combination of several ingredients of the StM calculation,
and a shortcoming of a particular ingredient of the StM may, at least partly,
be compensated by a further shortcoming of another ingredient.  

\subsubsection{$\chi^2/F$ from \ran\ data}
\label{sec:chi2an}
The cross section of the \ran\ reaction is governed by its significantly
negative $Q$-value of about $-9$\,MeV and the resulting phase space at
energies close above the threshold. The sensitivity to all parameters remains
limited, and for all 6720 combinations the $\chi^2/F$ remains within 4.7 and
16.3. The low sensitivity of the \ran\ cross section can also be seen in
Fig.~\ref{fig:reac_chi2}c). The lowest $\chi^2/F \approx 4.7$ is found for
A-OMP: Nolte potential \cite{Nol87}; N-OMP: Koning and Delaroche (TALYS
default); GSF: Kopecky-Uhl generalized Lorentzian, with normalized M1
strength; LD: microscopic, from Skyrme force, Hilaire's combinatorial
tables. For the \ran\ fit the overall $\chi^2/F$ increases significantly by
more than two orders of magnitude to 1265; i.e., because of the reduced
sensitivity of the \ran\ cross section, it is practically not possible to
constrain the parameters of the StM calculations from the \ran\ data.

\subsubsection{$\chi^2/F$ from \rap\ data}
\label{sec:chi2ap}
The \rap\ reaction dominates in the energy range under study, and by far the
most data points are available for this channel. It is not surprising that the
fit of the \rap\ data leads to a combination of parameters which is close to
the overall best-fit. Here we find A-OMP: DGG-3 with \Ndgg\ $= 0.9$; N-OMP:
Koning and Delaroche (TALYS default); GSF: Brink-Axel Lorentzian, M1
normalized; LD: microscopic, from Skyrme force, Hilaire's combinatorial
tables. A $\chi^2/F \approx 6.6$ is achieved for the \rap\ channel, and the
overall $\chi^2/F$ increases only slightly to about 11.4 (compared to 7.7 for
the best-fit). However, the \rap\ fit leads to a significant overestimation of
the \rag\ channel, leading to a $\chi^2/F \approx 54$ for the few data points
for the \rag\ channel.

\subsubsection{$\chi^2/F$ from \stot\ data}
\label{sec:chi2aa}
The two data points for \stot\ are best reproduced by the DGG-1 potential with
a $\chi^2/F \approx 2.9$ and an average deviation of only 3\,\%. The worst
description is obtained from the Nolte potential with a $\chi^2/F = 200$ and
an average deviation of 22\,\%. Most of the potentials under study reproduce
the 2 experimental data points with average deviations between 5 and 10\,\%,
and average deviations above 15\,\% are only found for the earlier Avrigeanu
potential \cite{Avr94} and the Nolte potential \cite{Nol87} which has been
optimized at much higher energies. 

\subsubsection{TALYS default}
\label{sec:TALYS_def}
In addition, Fig.~\ref{fig:reac_chi2} includes also the default TALYS
combination of A-OMP: Watanabe \cite{Wat58} (Note that this will change to
Avrigeanu \cite{Avr14} in the next versions.); N-OMP: Koning and Delaroche;
GSF: Kopecky-Uhl generalized Lorentzian, M1 normalized; LD: constant
temperature Fermi gas. The TALYS default calculation leads to an increased
$\chi^2/F \approx 309$ which results from a significant overestimation of the
\rap\ cross sections at low energies and a strong underestimation of the
\rag\ cross sections at all energies under study (see
Fig.~\ref{fig:reac_chi2}).

\section{Discussion}
\label{sec:disc}
One main task in nuclear astrophysics is the determination of reaction rates
\Nsv\ which are essentially defined by the cross sections at low energies. In
the present study we aim to use the above $\chi^2/F$ calculations to constrain
the \zniv \rag \geviii\ cross section for energies below the experimental
data. Before this can be done in the next Sect.~\ref{sec:extra}, the results
of the previous Sect.~\ref{sec:res} with $\chi^2/F \gg 1$ have to be
discussed in more detail. 

For the following discussion let us first assume that the statistical model is
valid for \zniv\ + \al , and at least one hypothetical and a priori unknown
combination of the almost 7000 combinations of input parameters is able to
reproduce the experimental cross sections. Under these assumptions one should
find that this hypothetical best-fit combination reproduces the experimental
data with $\chi^2/F \lesssim 1$. For dominating statistical uncertainties of
the experimental data, $\chi^2/F \approx 1.0$ should be found. For dominating
systematic uncertainties, even cases with $\chi^2/F \ll 1$ may be
found. Typical systematic uncertainties from charge integration, target
thickness, or detector efficiency affect all data points in the same
way (except \stot\ from elastic scattering). Thus, for dominating systematic
uncertainties, a variation of the absolute normalization of the experimental
data within their common systematic uncertainty should lead to an almost
perfect agreement between the hypothetical best-fit combination and the
normalized experimental data with $\chi^2/F \ll 1$.

In reality, the experimental data points are affected by both, statistical and
systematic, uncertainties. For most of the data points, the systematic
uncertainty is dominating; only for the weak \rag\ channel and for low
energies or energies close above the respective thresholds, the statistical
uncertainty is dominating \cite{Gyu12}. Thus, we have varied the absolute
normalization $N_{\rm{exp}}$ of the \rag , \ran , and \rap\ cross sections
within a range of $N_{\rm{exp}} = 0.7 - 1.3$ which corresponds to about 3
times the systematic uncertainty of the data of about 10\,\% \cite{Gyu12}. A
smooth variation of the $\chi^2/F$ with the normalization factor
$N_{\rm{exp}}$ was found with a minimum of $\chi^2/F = 7.59$ for $N_{\rm{exp}}
= 1.05$, compared to $\chi^2/F = 7.74$ for the original data ($N_{\rm{exp}} =
1.0$). Consequently, among the almost 7000 combinations of parameters for the
StM calculations, there is no combination with $\chi^2/F \ll 1$, i.e., none of
the almost 7000 combinations is able to reproduce the experimental
data. 

Strictly speaking, this means that either all almost 7000 combinations
of input parameters are incompatible with the experimental results, or the
chosen StM is inappropriate for the present case. However, neither a better
model is available for the calculation of the \zniv\ + \al\ reaction cross
sections, nor better parameterizations of the ingredients of the StM are
available; this holds in particular for the A-OMPs under study which govern
the theoretical uncertainties of the calculated low-energy cross sections.
Nevertheless, low-energy cross sections have to be calculated, and their
uncertainties have to be estimated, to provide an astrophysical reaction rate
\Nsv\ with a reasonable error bar. Therefore, the following considerations
were used to obtain a realistic constraint for the low-energy \rag\ cross
section.

For dominating statistical uncertainties of the experimental data under study,
the uncertainty of a fit parameter is calculated from the increase of $\chi^2$
by 1. Contrary, a dominating systematic uncertainty leads to a larger
uncertainty for fit parameters because systematic uncertainties from many data
points do on average not cancel each other. Thus, an increase of $\chi^2$ for
each data point by 1 should be used in the latter case; i.e., an increase of
$\chi^2/F$ by 1 indicates the uncertainty of a fit parameter in the case of
dominating systematic uncertainties. However, both
above criteria for $\chi^2$ for statistical uncertainties and for $\chi^2/F$
for systematic uncertainties are only robust if $\chi^2/F \lesssim 1$ is
achieved; but this was not possible in the present study of \al -induced
reaction cross sections of \zniv . Consequently, the available experimental
data for \zniv\ cannot provide a strict mathematical constraint for
\rag\ cross section at lower energies. 

A reasonable criterion for the allowed range of $\chi^2/F$ has to be chosen to
find a rather reliable constraint of the \rag\ cross section at low
energies. It should be pointed out here that this is not a new problem of the
present study. The problem becomes only very obvious here because we attempt
to provide a $\chi^2$-based constraint of the \rag\ cross section at low
energies. In many previous papers the best combination of parameters was
simply derived ``by eye'' from the comparison of experimental excitation
functions with theoretical predictions using a more or less broad range of
input parameters and/or computer codes for the StM. Then, often this best
combination is simply used to calculate astrophysical reaction rates (without
further discussion of $\chi^2$).

Following the above discussion, we have decided to use the following criterion
for the allowed variation of $\chi^2$. The best-fit combination of parameters
reaches $\chi^2/F \approx 7.7$, corresponding to an average deviation of about
30\,\% from the experimental data. For experimental uncertainties of the order
of 10\,\%, this means that we find an average deviation of almost
$3\sigma$. For the uncertainty determination of the low-energy \rag\ data, we
now accept all combinations of parameters with $\chi^2/F \le 15$. This
corresponds to an average $\approx 4\sigma$ deviation instead of the best-fit
$\approx 3\sigma$ deviation; i.e., we allow for an increase of the average
deviation of each data point by $1\sigma$, and the resulting parameter space
should describe the data with average deviations of less than about 40\,\%.

\subsection{Extrapolation to low energies}
\label{sec:extra}
In the following we restrict ourselves here to estimate the \rag\ cross
section at two energies below the lowest experimental point at about 8\,MeV.
The first energy is chosen from the most effective energy at $T_9 = 2.5$,
i.e.\ a temperature of $T = 2.5 \times 10^9$\,K which is typical for the
astrophysical \ppro . From the standard formulae
which are based on an energy-independent astrophysical \sfact , the most
effective energy at this temperature is $E_{\rm{eff}} \approx 5.36$\,MeV. In
practice, the assumption of a constant \sfact\ is not realistic for heavy
nuclei, and in most cases the effective energy is slightly shifted towards
lower energies \cite{Rau10}.

The second energy is taken as $E = 3.95$\,MeV which is slightly below the
\rap\ threshold and far below the \ran\ threshold. At this low energy the
\rag\ cross section depends almost exclusively on the chosen A-OMP.

\begin{figure}[htb]
\includegraphics[width=0.9\columnwidth,clip=]{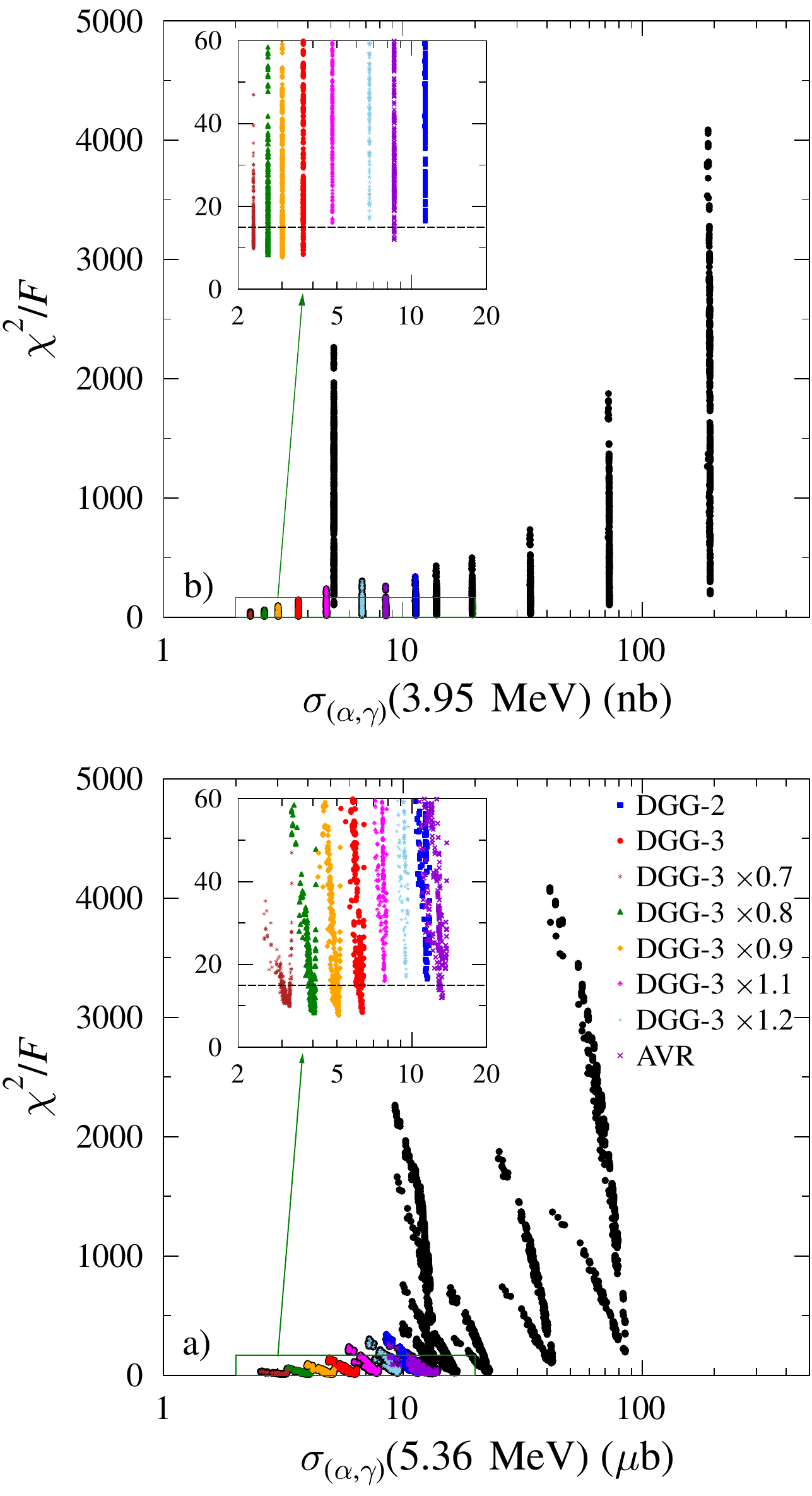}
\caption{
\label{fig:chi2}
(Color online) 
$\chi^2/F$ as a function of the \rag\ cross section at 5.36\,MeV (lower part;
corresponding to the effective energy at $T_9 = 2.5$) and at 3.95\,MeV (upper
part). Huge variations between about 8 and 5000 are found for
$\chi^2/F$. The insets show all calculations with small $\chi^2/F < 60$. The
chosen criterion $\chi^2/F \le 15$ is indicated by a horizontal dashed
lines. Further discussion see text.
}
\end{figure}

For the almost 7000 combinations of parameters, the calculated \rag\ cross
sections at 5.36\,MeV vary between 2.5\,$\mu$b and
85\,$\mu$b. Fig.~\ref{fig:chi2}a) shows that the corresponding $\chi^2/F$ vary
between about 8 and almost 5000. The inset shows the calculations with
$\chi^2/F < 60$. Here it becomes clearly visible that the $\chi^2/F$ are
grouped according to the chosen A-OMP; e.g., the DGG-3 potential (with
\Ndgg\ $ = 1.0$, i.e.\ without
further adjustment of the depth of the real part) favors \rag\ cross sections
between 4.9 and 6.5\,$\mu$b (with smallest $\chi^2/F$ for $\sigma \approx
6.3$\,$\mu$b), and the AVR potential favors cross sections between 8.7\,$\mu$b
and 13.8\,$\mu$b (with smallest $\chi^2/F$ for $\sigma \approx 13.2$\,$\mu$b).
Adopting the above criterion of $\chi^2/F \le 15$, we find $2.9\,\mu{\rm{b}}
\le \sigma_{(\alpha,\gamma)} \le 13.3\,\mu{\rm{b}}$ at $E = 5.36$\,MeV. Thus,
the chosen criterion $\chi^2/F \le 15$ restricts the \rag\ cross section to
$\sigma \approx 6$\,$\mu$b with an uncertainty of a factor of two, whereas the
range of all calculations was between 2.5\,$\mu$b and 85\,$\mu$b.

The same procedure is repeated for the lower energy of 3.95\,MeV (see
Fig.~\ref{fig:chi2}, upper part b). Here the predictions vary between 2.3\,nb
and 192\,nb, i.e.\ the predictions cover almost two orders of
magnitude. Applying the criterion $\chi^2/F \le 15$ restricts the \rag\ cross
section to $2.3\,{\rm{nb}} \le \sigma_{(\alpha,\gamma)} \le 8.5\,{\rm{nb}}$ at
$E = 3.95$\,MeV, i.e.\ the \rag\ cross section is $\sigma \approx 4.4$\,nb,
again with an uncertainty of about a factor of two. Combinations of parameters
which lead to much higher cross sections, are excluded by the chosen $\chi^2$
criterion.

As expected, the calculated cross section of the \rag\ reaction depends almost
exclusively on the A-OMP at the lower energy of 3.95\,MeV. This is reflected
by the strictly vertical grouping of the different A-OMPs in
Fig.~\ref{fig:chi2}b). At the slightly higher energy of 5.36\,MeV, a grouping
according to the A-OMPs is also observed. However, because the \rag\ cross
section is not only sensitive to the A-OMP, but also to other parameters, the
grouping is not strictly vertical here, see Fig.~\ref{fig:chi2}a).

Finally, it is interesting to see that the few-parameter ATOMKI-V1 potential
is only able to reach $\chi^2/F \approx 100$, but nevertheless it is able to
predict the low-energy \rag\ cross sections reasonably well. At the higher
energy of 5.36\,MeV the ATOMKI-V1 potential predicts \rag\ cross sections
between 9.3\,$\mu$b and 13.5\,$\mu$b, and at the lower energy of 3.95\,MeV the
\rag\ cross section from ATOMKI-V1 is 5.1\,nb.

\section{Conclusions}
\label{sec:conc}
Excitation functions of the cross sections of the \zniv \rag \geviii , \zniv
\ran \gevii , and \zniv \rap \gavii\ reactions and the total reaction cross
section \stot\ have been analyzed using the statistical model and a
$\chi^2$-based assessment of the underlying parameters. A best fit to the
experimental data at low energies \cite{Orn16,Gyu12} shows $\chi^2/F \approx
7.7$ and an average deviation factor of about \fdev\ $\approx 1.3$ from all
experimental data.

The complete parameter space of the TALYS code was investigated using almost
7000 combinations of the \al -nucleus OMPs, nucleon-nucleus OMPs, gamma-ray
strength functions, and level densities. As the most important ingredient of
these StM calculations, the \al -nucleus potential was identified. This fact
can be derived from the behavior of the $\chi^2/F$. The best-fit is obtained
by a modification of the third version of the Demetriou {\it et
  al.}\ \cite{Dem02} potential where the real part is scaled by a factor of
\Ndgg\ $ = 0.9$. This best-fit result is still far from $\chi^2/F \approx
1$. A reduction of $\chi^2$ should be achievable from further improvements of
the \al -nucleus OMP at low energies. $\chi^2/F \approx 1$ will probably not
be reachable because of non-statistical fluctuations of the reaction cross
sections, in particular for the \zniv \rap \gavii\ reaction at low energies.

From the range of $\chi^2/F$ and the corresponding variation of the
\rag\ cross section at lower energies, an uncertainty of about a factor of two
is estimated for the astrophysical reaction rate of the \zniv \rag
\geviii\ reaction and its cross section at very low energies. The uncertainty
results from all reasonable fits with $\chi^2/F \le 15$ and their
extrapolations down towards astrophysically relevant energies where no
experimental data for the \zniv \rag \geviii\ reaction are available. Improved
\rag\ data at lower energies could reduce the uncertainty of the
\rag\ reaction rate and help to further constrain the parameters for the
statistical model calculations.

\begin{acknowledgments}

This work is dedicated to Endre Somorjai on the occasion of his $80^{\rm{th}}$
birthday. We thank the members and collaborators of the Atomki Nuclear 
Astrophysics group - established by Endre Somorjai - for maintaining the
fruitful and pleasant working atmosphere over many years. 
This work was supported by OTKA (K108459 and K120666).
\end{acknowledgments}

\end{document}